# Open-World Darknet Traffic Recognition Under Leave-One-Service-Out Evaluation

Javeriah Saleem
*School of Computing, Mathematics and Engineering,*
*Charles Sturt University*
Wagga Wagga, Australia
jsaleem@csu.edu.au

Rafiqul Islam
*School of Computing, Mathematics and Engineering*
*Charles Sturt University*
Albury, Australia
mislam@csu.edu.au

Md Zahidul Islam
*School of Computing, Mathematics and Engineering*
*Charles Sturt University*
Bathurst, Australia
zislam@csu.edu.au

***Abstract*— Darknet traffic recognition is critical for cyber threat intelligence, as anonymity networks are often used to conceal malicious activity. However, most existing studies rely on closed-world evaluation, assuming all service categories are known during training and testing, which is unrealistic in real-world environments. This paper presents an open-world darknet traffic classification framework using leave-one-service-out evaluation and uncertainty-aware classification with Random Forest and XGBoost models. Experimental results demonstrate significant performance degradation when transitioning from closed-world to open-world settings, demonstrating that closed-world evaluation substantially overestimates deployment robustness. For example, XGBoost Macro-F1 decreases from 88.8% to 46.1% in the I2P environment, while Random Forest performance drops from 87.4% to 45.7%. Although uncertainty-based rejection slightly improves robustness, strong behavioral similarity between known and unknown services leads to frequent misclassification. Semantic absorption analysis further shows that FreeNet video traffic is classified as browsing traffic with an 88.1% assignment rate, while I2P peer-to-peer traffic is absorbed into FTP-related behavior with an 83.4% assignment rate. The findings demonstrate that behavioral overlap remains a major challenge for reliable open-world darknet traffic classification.**

## I. Introduction

The increasing adoption of encrypted communication and anonymity-preserving technologies has significantly complicated the analysis of modern network traffic. Anonymity networks such as Tor, I2P, FreeNet, and ZeroNet are widely used to protect user privacy through layered encryption and decentralized routing mechanisms [1]. However, these platforms are also exploited for cybercrime marketplaces, malware communication, illegal content distribution, and covert data exchange, making reliable darknet traffic classification an important problem in cybersecurity and cyber threat intelligence [2].

Recent machine learning and deep learning approaches have achieved strong performance in encrypted traffic classification using statistical flow descriptors, temporal behavior, and hierarchical classification frameworks [3]. Existing studies demonstrate that encrypted traffic still contains measurable behavioral patterns that can be exploited for browser- and application-level identification despite payload encryption. However, most darknet traffic classification studies rely on a closed-world assumption, where all service categories are available during both training and testing. Although this setting simplifies experimental evaluation, it fails to represent realistic deployment environments where classifiers may encounter previously unseen services during inference. Consequently, models evaluated under closed-world conditions may substantially overestimate their real-world robustness [4].

Open-world recognition addresses this limitation by evaluating classifier behavior when unknown categories appear during inference [5]. In darknet environments, unknown services often exhibit structural and temporal characteristics similar to trained services, leading to strong behavioral overlap between known and unknown service traffic. In this work, behavioral overlap refers to structural and temporal similarity between known and unknown darknet services that results in high-confidence semantic misclassification. As a result, classifiers may confidently assign unknown service traffic to behaviorally related known-service categories rather than rejecting them as unseen traffic.

Despite its importance, open-world darknet traffic classification remains largely unexplored at the service level. Existing studies primarily focus on closed-world browser or application classification and provide limited analysis of how classifiers behave when previously unseen darknet services appear during inference.

Unlike existing studies that primarily evaluate closed-world classification accuracy, this work systematically investigates semantic absorption and confidence-overlap characteristics under service-level open-world evaluation on the Darknet-dataset 2020 [3]. The framework integrates training-only preprocessing, Random Forest feature selection, probability calibration, and uncertainty-based rejection using Random Forest (RF) and XGBoost (XGB) classifiers across Tor, I2P, FreeNet, and ZeroNet environments. The main contributions of this work are summarized as follows:

- An open-world darknet traffic classification framework based on leave-one-service-out evaluation is proposed for assessing classifier robustness against previously unseen services.
- The effectiveness of uncertainty-based rejection is systematically evaluated across multiple darknet browser environments under open-world conditions.
- Semantic confusion analysis is introduced to identify behavioral absorption patterns between unknown and known darknet services.
- A confidence-distribution analysis is performed to investigate the limitations of uncertainty-based rejection under conditions of behavioral overlap.

The remainder of this paper is organized as follows. Section II reviews related studies, Section III presents the proposed methodology, Section IV discusses the experimental results, and Section V concludes the paper.

## II. Related Studies

Darknet traffic classification has become an important research area due to the increasing use of anonymity networks such as Tor, I2P, FreeNet, and ZeroNet for both privacy-preserving communication and malicious cyber activities [4]. Since payload inspection is ineffective in encrypted environments, existing studies primarily rely on statistical flow descriptors, temporal behavior, and machine learning techniques for browser- and application-level identification [6]. Recent work has employed Random Forest, Support Vector Machines, XGBoost, and deep learning models, including convolutional and recurrent neural networks, while hierarchical frameworks have further improved browser- and service-level classification [1][3][7][8]. These studies demonstrate that encrypted traffic retains measurable behavioral characteristics that enable accurate classification despite encryption and routing obfuscation.

However, existing darknet traffic classification studies predominantly adopt a closed-world setting, where all service categories are assumed to be available during both training and testing. Although this simplifies evaluation, it does not reflect operational environments in which previously unseen services may appear during inference, potentially leading to overly optimistic estimates of deployment performance.

Open-world recognition addresses this limitation by enabling classifiers to reject previously unseen samples using uncertainty estimation, confidence thresholding, or out-of-distribution detection techniques [9][10][11]. Existing open-set recognition approaches, including OpenMax, have demonstrated promising performance across several application domains, whereas existing studies of encrypted traffic primarily focus on improving classification accuracy using conventional machine learning and deep learning models, providing limited insight into semantic confusion and confidence overlap when encountering unknown services. [10][12]. Consequently, the behavioral mechanisms underlying open-world classification failure remain insufficiently explored. To address this gap, this work investigates open-world darknet traffic classification using leave-one-service-out evaluation across Tor, I2P, FreeNet, and ZeroNet environments and analyzes how behavioral similarity influences uncertainty-based rejection performance.

## III. Methodology

This section presents the proposed open-world darknet traffic classification framework and the experimental methodology used to evaluate classifier robustness against previously unseen services. The framework integrates leave-one-service-out evaluation, training-only preprocessing, feature selection, probability calibration, and uncertainty-based rejection to investigate open-world classification behavior across multiple darknet browser environments. In addition to conventional performance evaluation, semantic confusion and confidence-distribution analyses are performed to examine the behavioral mechanisms underlying open-world classification failure.

### *A. Open-World Darknet Traffic Classification Framework*

This study proposes an open-world darknet traffic classification framework using leave-one-service-out evaluation. Unlike conventional closed-world evaluation, where all service categories are available during training and testing, the proposed framework excludes one service during training. The excluded service is introduced only during inference as unknown service traffic. This enables systematic evaluation of classifier robustness against previously unseen darknet services.

Browser-specific service environments are constructed from the Darknet-2020 dataset for Tor, I2P, FreeNet, and ZeroNet traffic [3]. For each experimental rotation, one service category is excluded from training and treated as unknown traffic, while the remaining services form the known-service training space.

A training-only preprocessing pipeline is employed to prevent train-test leakage. Following preprocessing, Random Forest feature importance analysis is used to select the top 50 discriminative features for each experimental setting. Random Forest (RF) and XGBoost (XGB) classifiers are then trained using the selected feature space. Probability calibration and validation-based threshold selection are further integrated to support uncertainty-based rejection under open-world conditions.

During inference, unknown service traffic is merged with known-service traffic and evaluated under both forced-classification and rejection settings. Finally, semantic confusion analysis, confidence-distribution analysis, and multi-metric robustness evaluation are performed to investigate the behavioral mechanisms responsible for open-world classification failure. Fig. 1 summarizes the complete workflow from feature preprocessing to uncertainty-based decision making under the LOSO protocol.

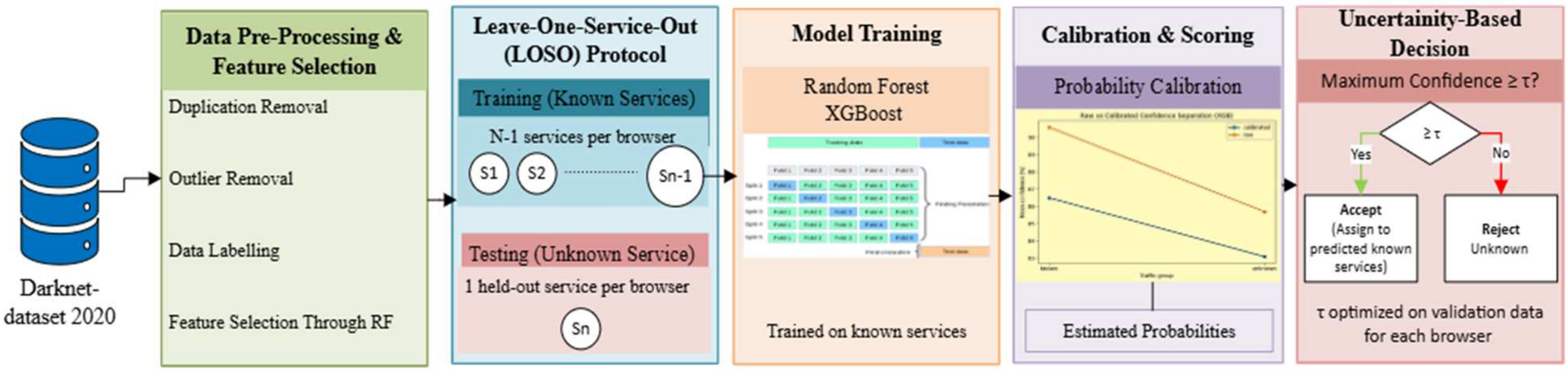

Fig. 1. Open-world darknet traffic classification framework using leave-one-service-out evaluation, feature selection, probability calibration, and uncertainty-based rejection.

### *B. Dataset and Browser-Specific Service Environments*

Experiments were conducted using the Darknet-dataset 2020, which contains encrypted traffic collected from four anonymity-network environments: Tor, I2P, FreeNet, and ZeroNet. Each browser environment contains multiple application-level service categories representing distinct communication behaviors.

To preserve browser-specific behavioral characteristics, each browser environment is evaluated independently. Table I summarizes the browser-specific service distributions used in this study.

For each browser environment, a leave-one-service-out (LOSO) evaluation is performed by treating one service as unknown service traffic while training the classifier on the remaining known services.

TABLE I. BROWSER-SPECIFIC SERVICES AND DISTRIBUTIONS.

| Browsers | Services | No. of Services |
|---|---|---|
| Tor | Audio, Browsing, Chat, Email, FTP, P2P, Video, VoIP | 8 |
| I2P | Browsing, Chat, Email, FTP, P2P | 5 |
| FreeNet | Browsing, Chat, Email, FTP, Video | 5 |
| ZeroNet | Audio, Browsing, Chat, Email, FTP, P2P, Video | 7 |

### C. Leave-One-Service-Out Open-World Protocol

Let $S = \{s_1, s_2, \dots, s_n\}$ denote the set of services within a browser environment. During each experimental rotation, one service $s_u \in S$ is excluded from training and treated as unknown service traffic. The remaining services form the known-service training space.

Known service traffic refers to traffic flows belonging to service categories included during training, whereas unknown service traffic represents held-out services introduced only during inference. This protocol simulates realistic deployment conditions where classifiers encounter previously unseen darknet services after deployment.

Each browser environment is evaluated independently to preserve service-specific behavioral relationships. During inference, unknown service traffic is merged with known service traffic and evaluated under two settings:

- Forced Classification: Every traffic instance is assigned to one of the known service categories.
- Uncertainty Rejection: Traffic instances with prediction confidence below a selected threshold are rejected as unknown. The uncertainty rejection mechanism is defined as:

$$\hat{y} = \begin{cases} arg\,max_i\, P(y_i \mid x), & max\, P(y_i \mid x) \geq \tau \\ Reject, & max\, P(y_i \mid x) < \tau \end{cases} \quad (1)$$

where $P(y_i \mid x)$represents the predicted probability for the class $y_i$, and $\tau$ denotes the uncertainty rejection threshold [13].

### D. Preprocessing and Feature Selection

A training-only preprocessing pipeline is employed to ensure leakage-safe evaluation. Median imputation, percentile clipping, and z-score normalization are fitted exclusively on the training partition and subsequently applied to the validation and test data.

Following preprocessing, Random Forest feature importance is used to rank traffic descriptors, and the top 50 features are retained for model training. Preliminary validation experiments showed that this feature subset provides a stable balance between dimensionality reduction and open-world classification performance. Random Forest (RF) and XGBoost (XGB) are then trained using the selected features. These classifiers were selected because they represent widely adopted and highly competitive tree-based approaches for encrypted traffic classification. This enables the proposed framework to evaluate the impact of open-world service conditions independently of classifier-specific architectural differences. In addition, the combination of feature selection, tree-based learning, and offline probability calibration results in a computationally efficient framework suitable for practical deployment.

For each leave-one-service-out rotation, the dataset is divided into training, validation, and test partitions using a 60/20/20 split. The validation partition is used for probability calibration and rejection-threshold optimization, while the test partition is reserved exclusively for final open-world evaluation. Two probability estimation settings are considered:

- Raw Probabilities: direct classifier probability outputs.
- Calibrated Probabilities: validation-calibrated probability estimates.

Open-world performance is evaluated using Accuracy, Precision, Recall, and Macro-F1 metrics under both forced classification and uncertainty rejection settings. All classifiers were evaluated using identical leave-one-service-out rotations across browser environments to ensure consistent comparative evaluation.

## IV. RESULTS AND DISCUSSION

This section presents the experimental results obtained under closed-world and open-world evaluation settings across Tor, I2P, FreeNet, and ZeroNet environments. The analysis focuses on classifier robustness, uncertainty-based rejection performance, semantic confusion behavior, and confidence-distribution characteristics when previously unseen service traffic is introduced during inference. Results are reported using Accuracy, Precision, Recall, and Macro-F1 metrics, along with behavioral analysis, to provide a comprehensive understanding of the performance of open-world darknet traffic classification.

### A. Closed-World Inflation under Open-World Evaluation

This experiment evaluates how classifier performance changes as it transitions from conventional closed-world evaluation to open-world service-level recognition. Table II summarizes the Macro-F1 performance of Random Forest (RF) and XGBoost (XGB) under closed-world classification, open-world forced classification, and uncertainty-based rejection settings.

The results show substantial degradation across all browser environments once unknown service traffic is introduced during inference. Under closed-world settings, RF and XGB achieve relatively strong classification performance, particularly within FreeNet and ZeroNet environments. However, open-world forced classification results in a significant performance collapse, indicating that conventional closed-world evaluation substantially overestimates deployment robustness.

The largest degradation is observed within the I2P environment, where XGB performance decreases from 88.8% to 46.1% Macro-F1 after introducing unknown service traffic.

Similar behavior is observed for RF, demonstrating that classifiers trained exclusively on known services struggle to generalize when previously unseen darknet services appear during inference.

Uncertainty rejection partially improves performance in selected environments, particularly within Tor, where XGB recovery increases by 8.5%. However, the improvement remains inconsistent across browser environments, suggesting that uncertainty estimation alone is insufficient for reliably separating known and unknown darknet services. Fig. 2 further illustrates the strong performance inflation caused by closed-world evaluation.

TABLE II. CLOSED-WORLD AND OPEN-WORLD MACRO-F1 COMPARISON UNDER UNCERTAINTY REJECTION.

| **Browser** | **RF Closed (%)** | **RF Forced (%)** | **RF Rejection (%)** | **RF Recovery** | **XGB Closed (%)** | **XGB Forced (%)** | **XGB Rejection (%)** | **XGB Recovery** |
|---|---|---|---|---|---|---|---|---|
| Tor | 76.2 | 54.1 | 57.4 | +3.3 | 78.5 | 56.4 | 64.9 | +8.5 |
| I2P | 87.4 | 45.7 | 48.0 | +2.3 | 88.8 | 46.1 | 46.1 | +0.0 |
| FreeNet | 96.0 | 61.9 | 62.7 | +0.8 | 95.9 | 63.6 | 66.3 | +2.7 |
| ZeroNet | 87.6 | 62.3 | 64.1 | +1.8 | 89.8 | 66.4 | 69.6 | +3.2 |

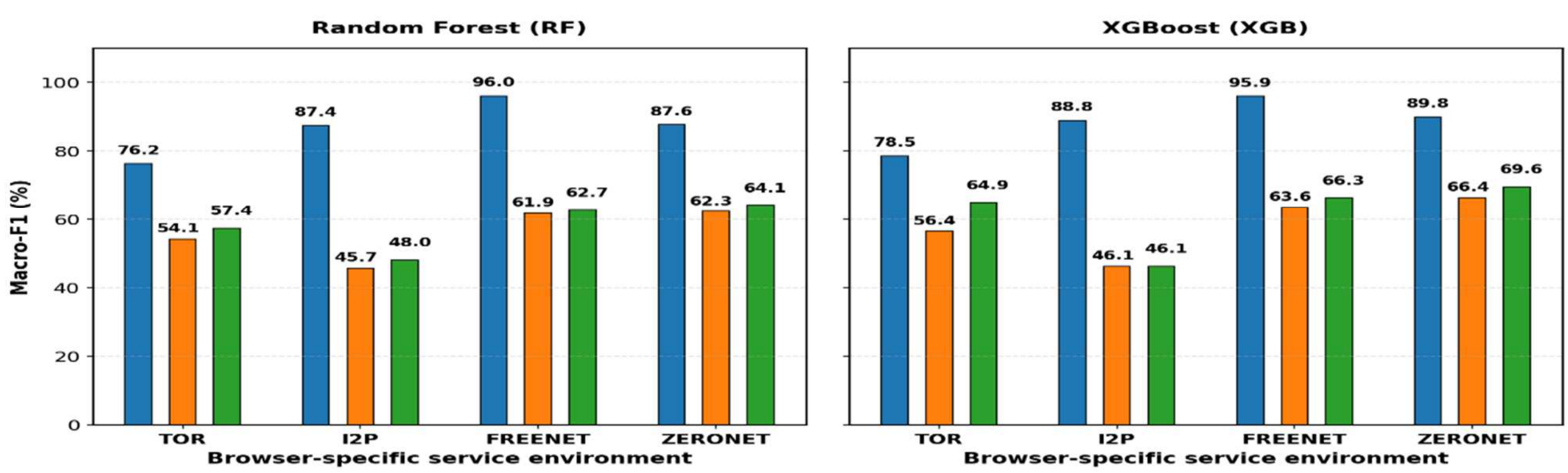

Fig. 2. Closed-world versus open-world Macro-F1 performance across browser environments.

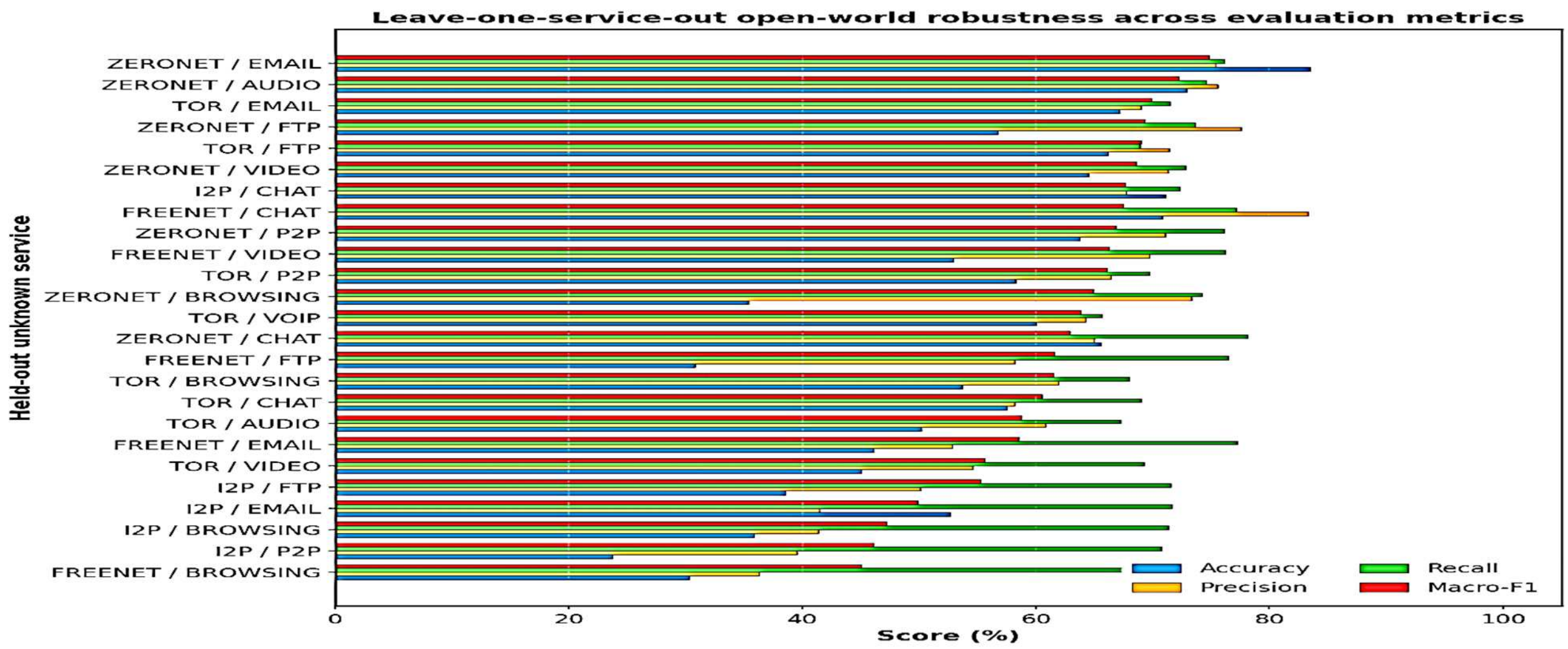

Fig. 3. Multi-metric open-world robustness analysis across leave-one-service-out browser environments.

## B. *Open-World Performance Behavior Across Browsers*

To further evaluate classifier robustness, additional experiments were conducted using Accuracy, Precision, Recall, and Macro-F1 metrics across all leave-one-service-out rotations. Fig. 3 summarizes the multi-metric robustness behavior across browser environments.

The results show that Macro-F1 degrades more than Accuracy does, indicating that open-world classification failures disproportionately affect behaviorally overlapping services. This behavior is particularly visible within the Tor and I2P environments, where Recall and Macro-F1 decrease substantially after introducing unknown service traffic.

Although FreeNet maintains comparatively stable aggregate metrics, subsequent semantic analysis reveals that unknown services are frequently absorbed into behaviorally similar known services with high confidence. Consequently, high Accuracy values do not necessarily indicate successful handling of unknown services. In contrast, Tor exhibits comparatively larger metric instability, suggesting partial separation between known and unknown service traffic distributions.

Overall, the multi-metric analysis demonstrates that conventional performance metrics alone cannot fully capture open-world classification behavior and should be complemented with behavioral analysis.

## C. *Semantic Absorption of Unknown Services*

To investigate the behavioral mechanisms underlying open-world classification failure, a semantic confusion analysis was performed to identify how unknown service traffic is absorbed into known service categories. Table III summarizes the dominant semantic absorption patterns observed across browser environments.

The results demonstrate that unknown service traffic is not randomly distributed across known classes. Instead, unseen services are consistently absorbed into behaviorally related known services. For example, FreeNet video traffic is overwhelmingly classified as browsing traffic, while I2P peer-to-peer traffic is strongly absorbed into FTP-related behavior.

These findings indicate that open-world failure is primarily driven by behavioral similarity rather than simple classifier uncertainty. When unknown services exhibit structural and temporal characteristics similar to those of trained services, classifiers confidently assign them to semantically related known categories rather than rejecting them as unknown traffic. Fig. 4 further illustrates these semantic absorption patterns through the semantic confusion heatmap.

Table III demonstrates that open-world failure is strongly associated with concentrated semantic absorption behavior and high levels of confidence in unknown services. Browser environments such as I2P and FreeNet exhibit both high dominant misclassification ratios and minimal browser-level rejection recovery, indicating that unknown services are consistently absorbed into semantically related known-service categories despite uncertainty-based rejection. In contrast, Tor exhibits comparatively weaker absorption behavior and lower confidence values, enabling stronger recovery under open-world evaluation.

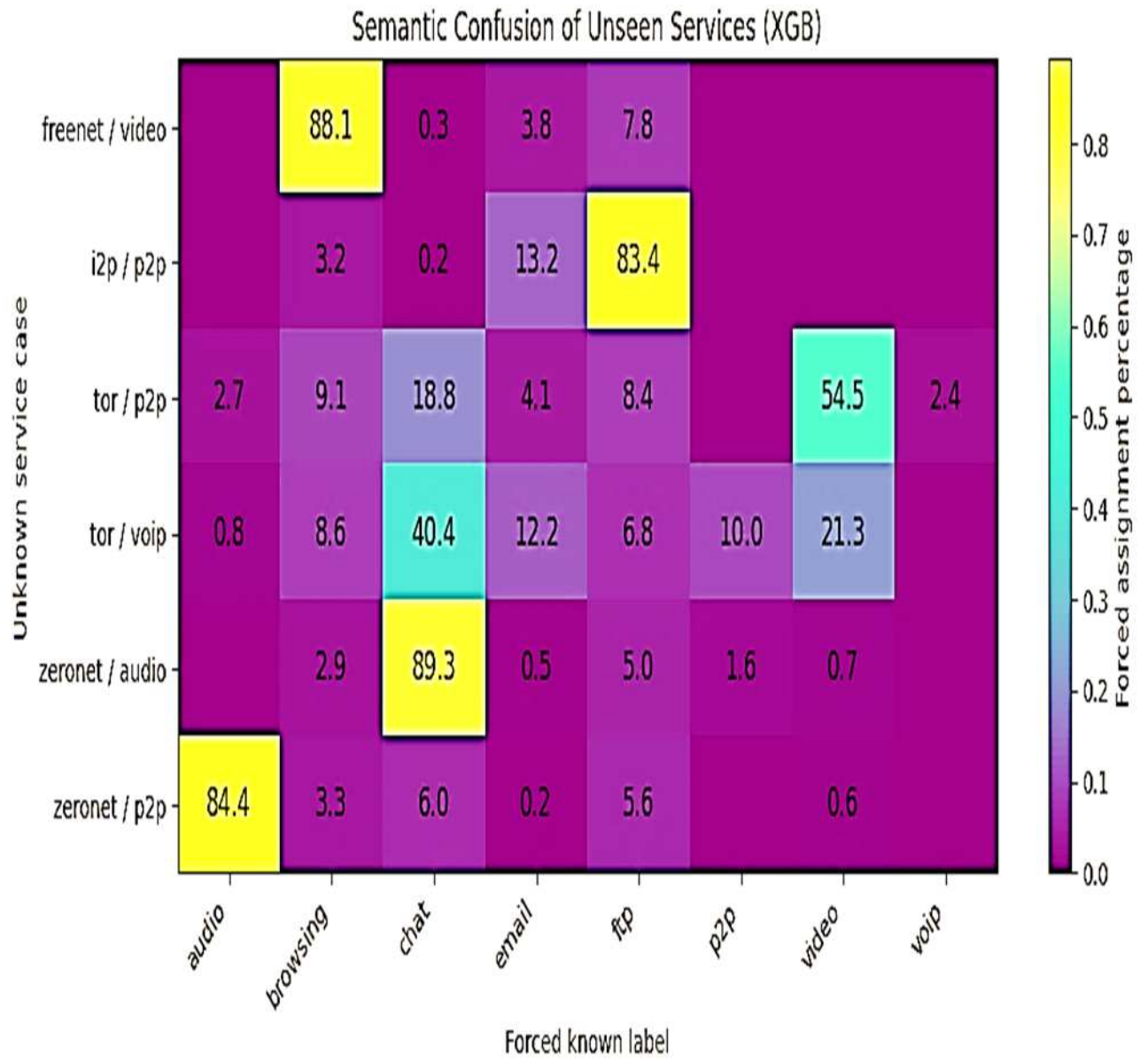

Fig. 4. Semantic confusion heatmap of unknown-service absorption behavior.

TABLE III. SEMANTIC ABSORPTION BEHAVIOR OF UNKNOWN SERVICES.

| Browser | Unknown Service | Assigned Known-Service | Misclassification Ratio (%) | Mean Confidence (%) | Browser-Level Recovery (%) |
|---|---|---|---|---|---|
| ZeroNet | Audio | Chat | 89.27 | 92.39 | +3.20 |
| FreeNet | Video | Browsing | 88.11 | 97.95 | +2.70 |
| ZeroNet | P2P | Audio | 84.36 | 92.77 | +3.20 |
| I2P | P2P | FTP | 83.44 | 92.93 | +0.00 |
| Tor | P2P | Video | 54.52 | 67.56 | +8.50 |
| Tor | VoIP | Chat | 40.37 | 67.83 | +8.50 |

## D. *Confidence Overlap and Rejection Failure*

To investigate the limitations of uncertainty-based rejection, confidence-distribution analysis was performed for both known and unknown service traffic across all browser environments. Table IV summarizes the mean, median, and standard deviation of classifier confidence scores.

Table IV shows significant overlap in confidence between known and unknown service traffic across multiple browser environments. FreeNet and I2P exhibit nearly identical confidence distributions for both known and unknown traffic, indicating that unseen services receive highly confident predictions despite being absent during training. Similar behavior is also observed in ZeroNet, where unknown service traffic maintains confidence levels close to those of trained services.

TABLE IV. CONFIDENCE STATISTICS FOR KNOWN AND UNKNOWN SERVICES

| **Browser** | **Traffic Type** | **Mean Confidence (%)** | **Median Confidence (%)** | **Std (%)** |
|---|---|---|---|---|
| Tor | Known Service | 81.22 | 96.84 | 24.57 |
| Tor | Unknown Service | 67.66 | 68.05 | 24.24 |
| I2P | Known Service | 92.30 | 99.78 | 17.09 |
| I2P | Unknown Service | 92.93 | 99.82 | 15.34 |
| FreeNet | Known Service | 97.75 | 99.96 | 9.81 |
| FreeNet | Unknown Service | 97.95 | 99.95 | 8.59 |
| ZeroNet | Known Service | 94.43 | 99.87 | 15.01 |
| ZeroNet | Unknown Service | 92.63 | 99.74 | 17.58 |

In contrast, Tor demonstrates partial confidence separation between known and unknown service traffic. Unknown Tor services yield lower mean confidence values and a broader distribution, enabling uncertainty-based rejection to achieve comparatively stronger performance in open-world evaluation.

Fig. 5 further visualizes the confidence distributions within Tor and FreeNet environments. Tor exhibits partial separation between known and unknown service traffic distributions, whereas FreeNet shows near-complete overlap. These findings demonstrate that uncertainty-based rejection is limited primarily because unknown services often exhibit behavioral characteristics highly similar to those of trained services, leading to overly confident misclassification rather than rejection.

These findings suggest that deploying conventional closed-world darknet classifiers in operational environments may lead to overconfident predictions for previously unseen services, thereby reducing reliability in dynamic network environments.

Overall, the results confirm that behavioral overlap represents a major challenge for open-world darknet traffic classification. While uncertainty-based rejection provides partial improvement in selected environments, classifiers remain vulnerable when unknown services closely resemble known service behavior.

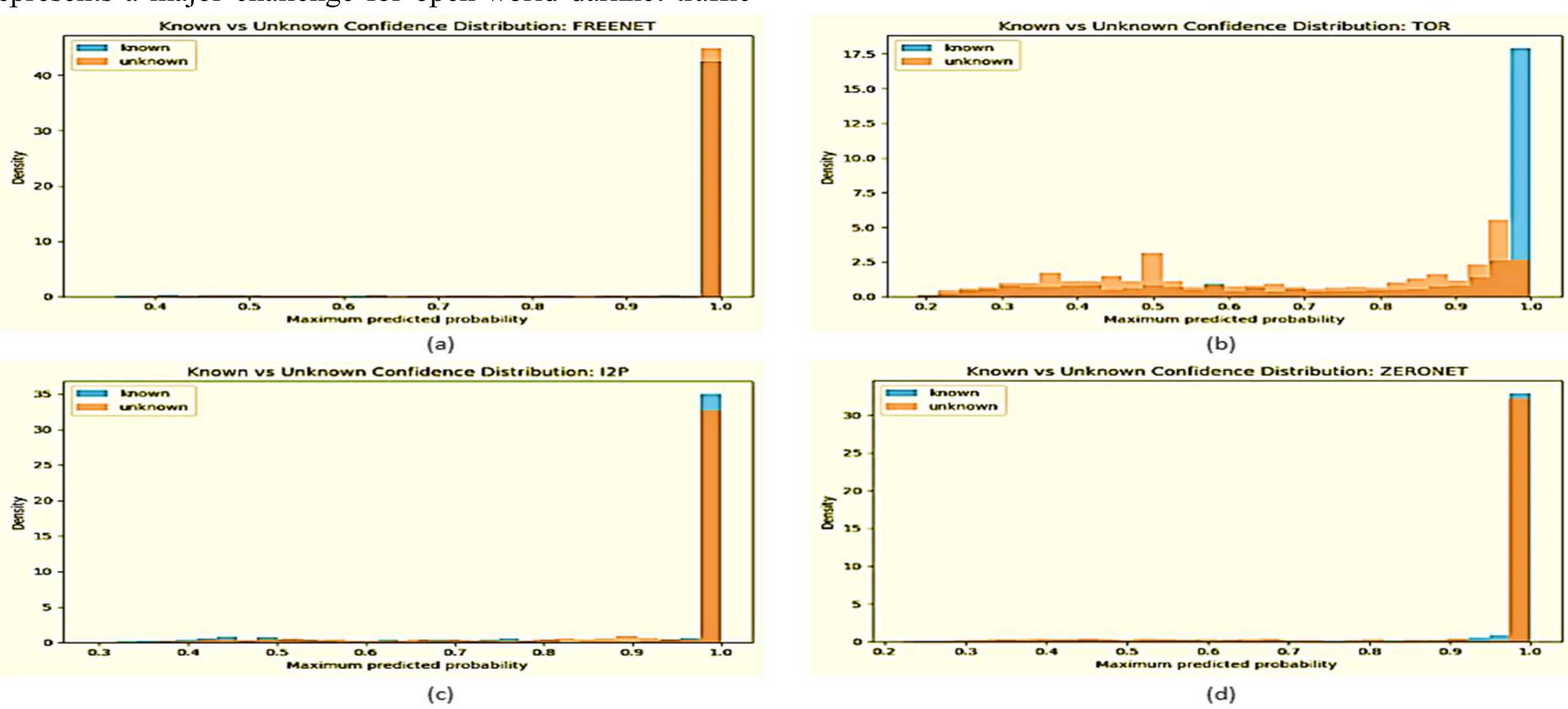

Fig. 5. Maximum confidence distributions for known and unknown service traffic across browser environments.

## V. CONCLUSION

This paper presented an open-world darknet traffic classification framework based on leave-one-service-out evaluation to assess classifier robustness against previously unseen service traffic across Tor, I2P, FreeNet, and ZeroNet environments. Unlike conventional closed-world evaluation, the proposed framework systematically examines classifier behavior when encountering unknown services during inference. Experimental results demonstrated substantial performance degradation under open-world conditions, indicating that closed-world evaluation significantly overestimates deployment robustness. Although uncertainty-based rejection improved performance in selected browser environments, semantic absorption and confidence-distribution analyses revealed that strong behavioral overlap between known and unknown service traffic often leads to highly confident misclassifications, thereby limiting the effectiveness of confidence-based rejection.

The current study is limited to a single publicly available darknet dataset, browser-specific service environments, and fixed confidence-threshold rejection settings. Although representative open-set recognition methods, such as OpenMax, were not evaluated because they rely on fundamentally different deep feature representations, their integration and comparative evaluation constitute an important direction for future work. Future research will further investigate cross-dataset generalization, advanced open-set recognition techniques, and out-of-distribution detection methods to improve the robustness and deployment reliability of open-world darknet traffic classification systems.